\documentclass[journal]{IEEEtran}
\ifCLASSINFOpdf
\else
\fi
\usepackage{hhline}
 \usepackage{amssymb,amsmath,epsfig,cite,url,multicol,multirow}
\usepackage{hyperref}
\usepackage{graphicx}
\usepackage{float}
\newcommand{\ra}[1]{\renewcommand{\arraystretch}{#1}}

\begin{document}
%
% paper title
% can use linebreaks \\ within to get better formatting as desired
\title {A Deep Learning Algorithm for Objective Assessment of Hypernasality in Children with Cleft
Palate }

% Objective assessment of hypernasality in children with cleft palate via deep model transfer

%
%
% author names and IEEE memberships
% note positions of commas and nonbreaking spaces ( ~ ) LaTeX will not break
% a structure at a ~ so this keeps an author's name from being broken across
% two lines.
% use \thanks{} to gain access to the first footnote area
% a separate \thanks must be used for each paragraph as LaTeX2e's \thanks
% was not built to handle multiple paragraphs
%

\author{Vikram C. Mathad, Nancy Scherer, Kathy Chapman, Julie M. Liss, and Visar Berisha  
\thanks{Vikram C. Mathad is with the Department of Speech \& Hearing Sciences, Arizona State University, Tempe, AZ-85281, Email: vchikkaw@asu.edu.  Nancy Scherer is with the Department of Speech \& Hearing Sciences, Arizona State University, Tempe, AZ-85281, Email: nancy.scherer@asu.edu.  Kathy Chapman is with the Department of Communication Sciences and Disorders, University of Utah, Salt Lake City, UT-84112, Email:kathy.chapman@health.utah.edu. Julie M. Liss is with the Department of Speech \& Hearing Sciences, Arizona State University, Tempe, AZ-85281, Email: julie.liss@asu.edu. Visar Berisha is with the College of Health Solutions, and School of Electrical, Computer, \& Energy Engineering, Arizona State University, Tempe, AZ-85281, Email: visar@asu.edu. This work is funded in part by NIH grant NIDCR DE026252}}

\maketitle

\begin{abstract}

 Objectives: Evaluation of  hypernasality requires extensive perceptual training by clinicians and extending this training on a large scale internationally is untenable; this compounds the health disparities that already exist among children with cleft. In this work, we present the objective hypernasality measure (OHM), a speech analytics algorithm that automatically measures hypernasality in speech, and validate it relative to a group of trained clinicians.  Methods: We trained a deep neural network (DNN) on approximately 100 hours of a publicly-available healthy speech corpus to detect the presence of nasal acoustic cues generated through the production of nasal consonants and nasalized phonemes in speech. Importantly, this model does not require any clinical data for training. The posterior probabilities of the deep learning model were aggregated at the sentence and speaker-levels to compute the OHM.  
  Results: The results showed that the OHM was significantly correlated  with the perceptual  hypernasality ratings in the Americleft database ({\it r=0.797, ~p$<$0.001}), and with the New Mexico Cleft Palate Center (NMCPC) database ({\it r=0.713,~p$<$0.001}).  In addition, we evaluated the relationship between the OHM and articulation errors; the sensitivity of the OHM in detecting the presence of very mild hypernasality; and establishing the internal reliability of the metric. Further, the performance of OHM was compared with a DNN regression algorithm directly trained on the hypernasal speech samples.  Significance: The results indicate that the OHM is able to rate the severity of hypernasality on par with Americleft-trained clinicians on this dataset.

\end{abstract}
%\end{IEEEbiography}

\section{Introduction}

Cleft palate (CP), with or without cleft lip, is a craniofacial anomaly and the most common birth disorder, with  1 in every 700  live births presenting with craniofacial clefts \cite{mossey2003global}. In healthy craniofacial development, the bilateral bony palatal shelves fuse horizontally at midline to create the roof of the mouth (hard palate) and provide points of muscular attachment for the soft palate (velum). These velar muscles, along with those in the upper pharynx, allow for modulation of the opening between the oral and nasal cavities (velopharyngeal port) during respiration, swallowing, and speaking. The failure of the palatal shelves to fuse at midline during embryological development (cleft) means there is no hard or soft palate and no separation between the oral and nasal cavities. The primary intervention involves surgical repair of the palatal cleft  to produce anatomical closure and to create the ability to modulate the velopharyngeal port aperture.  When velopharyngeal dysfunction (VPD) persists post primary palate surgery ~\cite{kummer2013cleft}, a secondary surgery (e.g. pharyngeal flap,  dynamic sphincter pharyngoplasty) is required. In the presence of VPD, the velopharyngeal port fails to close off the nasal tract completely during speech production for non-nasal sounds, and air and acoustic energy escape through the nasal cavity, resulting in reduced speech intelligibility. Twenty to thirty percent of children with clefts require a secondary surgery to rectify VPD~\cite{hardin2005speech} for the exclusive purpose of improving speech outcomes.

The inability to achieve adequate velopharyngeal closure during speech results in the percept of hypernasality, characterized by excessive nasal resonance due to passage of the vibrating column of air  through the nasal cavity (see supplementary material to listen to hypernasal speech). The perception of hypernasality in speech, secondary to VPD, is considered a primary outcome measure in CP as it drives decisions related to secondary surgery, speech therapy, and is an important determinant of long-term educational and social outcomes~\cite{chapman2011relationship,bell2017school}. As a result, it is considered a primary outcome by the American Cleft Palate-Craniofacial Association and the Cleft Palate Committee of the International Association of Logopedics and Phoniatrics~\cite{ref_CLP}.

 Instrumental methods such as a nasometer, magnetic resonance imaging, and cineradiography can be used to assess  VPD~\cite{bettens2014instrumental}.  These instruments require special training and expensive equipment; furthermore, they only show a moderate correlation with perceptual impressions of hypernasality. As a result, they are rarely used in clinical practice ~\cite{bettens2014instrumental}. Instead, clinicians rely on their perception of hypernasality to assess VPD.  Perception of hypernasality is a complex task that requires the clinician to infer, from the acoustic signal, the ratios of resonances across the pharyngeal, oral, and nasal cavities. The clinician then maps the perceived ratios to equal-interval or visual-analog scales of hypernasality. However, this percept is vulnerable to other co-modulating variables such as the words being spoken, the quality and loudness of the voice, audible turbulence and escape of air through the nose (nasal emission), and the idiosyncratic shape of an individual’s resonating cavities~\cite{baylis2015validity, yamashita2018reliability}. This results in a highly nonlinear mapping between the percept and the actual acoustic nasal resonance and considerable inter-rater and intra-rater variability in the assessment. Fundamentally, this limits the reliability and validity of the ratings obtained from untrained clinicians ~\cite{sweeney2008relationship}. The Americleft Speech Project was developed to address this dilemma by facilitating inter-center collaborations for speech outcomes research ~\cite{chapman2016americleft}. The first step included the development of a standardized protocol and calibration of craniofacial speech-language pathologists (SLPs) on perceptual ratings of hypernasality. Over the study period, recalibration was required to maintain high levels of inter-rater reliability. To date, only a small number of clinicians have participated in this program and applying this training on a large scale internationally is untenable.

\begin{figure*}[tbh]
                 
                   \centering
                   
 \includegraphics[height=50mm,width=\linewidth]{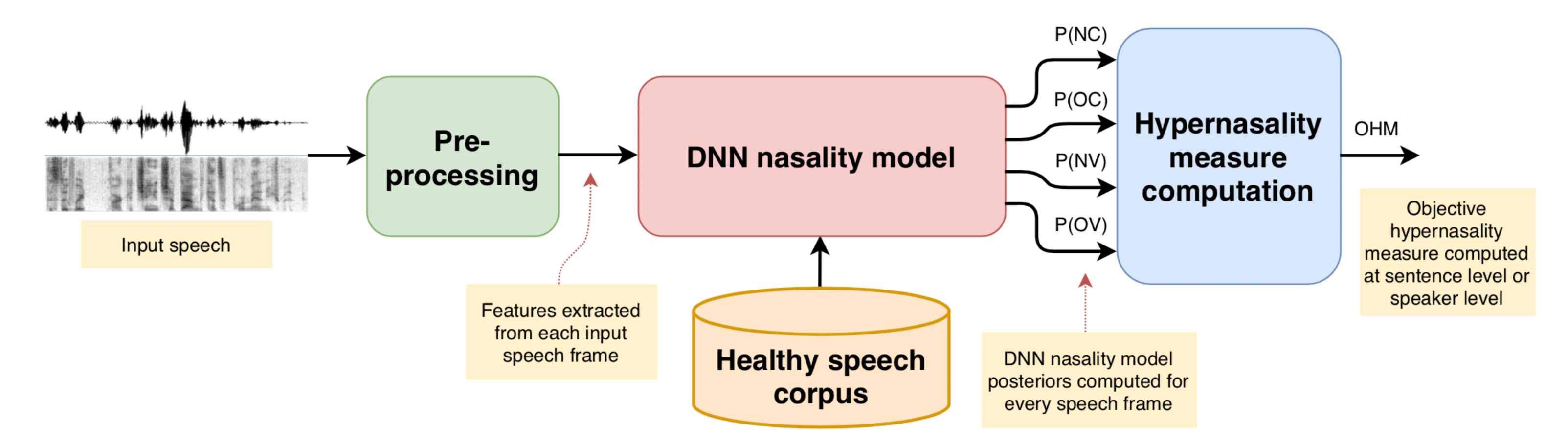}
                
  \caption{\label{block} Overview of the proposed approach for the hypernasality prediction. First, the input children's speech is pre-processed and passed through the pre-trained DNN Nasality model.  The DNN model posteriors are combined to form the objective hypernasality measure. }
               
\end{figure*}

In this paper, we present the objective hypernasality measure (OHM) to  assess hypernasality in CP speech and show that it tracks with the clinical perception of Americleft-trained SLPs. When clinicians make judgements of hypernasality, they focus on specific acoustic cues that are hallmarks of hypernasal speech. Similarly, we design an automatic assessment tool based on deep learning and demonstrate that the learned features from speech correlate with the clinical ratings of hypernasality. 

\subsection{Related work }
The development of speech technology-based system involves the extraction of acoustic features, which reflect abnormal nasal resonances present in the hypernasal speech.  Spectral measures, such as addition of extra nasal formant around 250 Hz, increased spectral flatness, reduced first formant amplitude, voice low-to-high tone ratio, and vowel space area have previously shown a correlation with the perceived hypernasality~\cite{kataoka1996spectral, vikram2016spectral, vijayalakshmi2007acoustic, lee2009evaluation, lee2006voice, nikitha2017hypernasality}. Acoustic features in combination with the machine learning algorithms have been used to develop automatic hypernasality assessment systems.   Mel-frequency cepstral coefficients (MFCCs), jitter, shimmer, vowel space area, wavlet transform based features have been used to train classifiers (e.g. support vector machines (SVMs), Gaussian mixture models (GMMS)) that detect hypernasal speech~\cite{lee2009evaluation, lee2006voice,golabbakhsh2017automatic,orozco2013nonlinear}.  Recently, convolutional neural network and recurrent neural networks have also been used for the same purpose~\cite{wang2019hypernasalitynet,wang2019automatic}. 

Most of these automatic algorithms  for detection of hypernasality were developed in a binary classification setting, i.e., healthy vs. hypernasal speech~\cite{lee2009evaluation, lee2006voice,golabbakhsh2017automatic,orozco2013nonlinear, wang2019automatic, wang2019hypernasalitynet}. This is inconsistent with  clinical practice, where clinicians require more fine-grained information (e.g. evaluation of hypernasality on a scale that ranges from normal to severe) for decision making. For example, a secondary surgery may only be required for treating moderate-severe hypernasal cases~\cite{kummer1996evaluation}.  There is a limited number of multi-class classification~\cite{dubey2019detection,he2014automatic,he2015automatic} and regression-based approaches for predicting hypernasality severity~\cite{mathad2020deep,saxon2019objective}.  These approaches rely on analysis of sustained phonations or segmented phonemes from utterances. This is limiting in two ways: (1) Sustained phonations don't capture phonetic context and don't provide a reliable percept of hypernasality. That's why clinicians prefer to use connected speech for reliable estimation of hypernasality~\cite{henningsson2008universal}. (2) For approaches that rely on connected speech, the phonetic segmentation was achieved either by manual marking or forced-alignment using orthogrophic transcriptions. However,  orthographic transcription is time consuming and the forced alignment procedure is prone to errors for children's speech~\cite{knowles2018examining}.

 Hypernasality estimation based on supervised machine learning requires large, labeled speech corpora.  Most of the existing speech-based hypernasality evaluation methods use speech samples and corresponding perceptual ratings to train machine learning models, including $k$-near neighborhood classifier~\cite{he2015automatic}, Gaussian mixture models~\cite{he2014automatic}, support vector machines~\cite{orozco2013nonlinear,dubey2019detection,golabbakhsh2017automatic, javid2020single}, and deep neural networks~\cite{wang2019automatic, wang2019hypernasalitynet,vikram2018estimation}.  The performance of these systems critically depends on the availability of clinical hypernasal speech databases that include speech samples from patients and corresponding hypernasality ratings from trained SLPs. However, development of a large hypernasal speech database  is difficult in practice due to the limited availability of patients' speech and the associated SLP clinical ratings. As a result, the models run the risk of overfitting to a particular database and rating scale.

\subsection{The proposed approach}
While large-scale databases of CP speech are untenable, healthy speech provides us with clues as to the acoustic manifestation of hypernasality. For example, the voiced sounds /M/ and /N/, and the sounds that precede and follow them, require opening of the velopharyngeal port to shunt the vibrating column of air through the nasal cavity. Thus, /M/ and /N/ are nasalized consonants (NC). Because the velum is a relatively sluggish articulator in comparison with the tongue and lips, the velopharyngeal port opens and closes more slowly, creating nasalization of vowels adjacent to the NCs, or nasalized vowels (NV). For example, the vowel /AE/ is nasalized in the word ``man''.  

This is in contrast to production of the oral consonants (OC), which involve closure of the velopharyngeal port to impound oral air pressure that creates a burst upon release of the articulatory closure (plosive). The voiced stop consonants, /B/ and /D/, and unvoiced stop consonants, /P/ and /T/, share the exact same places of articulation as /M/ and /N/, respectively, but are completely orally produced. This means that vowels (oral vowels, OV) adjacent to these OCs are also not nasalized, as in the /AE/ in the word ``cat''.  Since, the effect of nasalization is evident in the healthy speech, the acoustic manifestation of velopharyngeal port can be modelled using healthy speech corpus. Compared to CP speech, there are a large number of healthy speech corpora are available in the public domain. In our algorithm, we make use of a publicly available healthy speech corpus and train a nasality feature extraction model using only healthy speech. This results in an objective measure of hypernasality (OHM) that can be computed frame-by-frame and aggregated at the level of an utterance or speaker. 
  
An overview of the proposed algorithm for estimating the OHM is described in Fig.~\ref{block}. To learn the acoustic manifestation of the velopharyngeal port opening, we utilize 960 hours of speech from a publicly-available Librispeech corpus to train a deep neural network (DNN) model that classifies among nasalized consonants (NC), oral consonants (OC), nasalized vowels (NV), and oral vowels (OV). Training the DNN to classify among these classes forces it to learn the acoustic manifestation of an open velopharyngeal port. As a result, we refer to this DNN as the nasality model and use the four DNN posteriors of this model as ``features" for assessing the presence of nasalization in CP speech.  

 The input children's speech was pre-processed and we extracted the four posterior features using the pre-trained DNN nasality model. These features were combined to derive the OHM for each speech utterance. The details of the algorithm can be found in the Methods section. We established the construct validity of the OHM through several experiments using cleft speech samples and gold-standard clinical ratings from the Americleft project; then we evaluated the external validity of the model using data from the New Mexico cleft palate center (NMCPC) database. 
 
 \section{Databases}
 The details of the healthy speech corpus and the two CP speech databases are described below. 
  
\subsection{Healthy speech corpus}
 One hundred hours of healthy speech from the Librispeech database ({\it train-clean-100}) was used to train the DNN~\cite{panayotov2015librispeech}. The database contains English read speech samples recorded from 251 healthy adult speakers (125 male and 126 female). In addition to the speech samples, the database also contains orthographic transcriptions for each read sentence. A separate test set ({\it test-clean}) comprised of 5.4 hours of speech was used as a validation set. 
 
 \subsection{Americleft database} 
The Americleft database was collected as a part of the Americleft Speech Project at the University of Utah. The database consists of 60 children with CP (37 boys and 23 girls) of average age $6.276\pm0.676$ years. The control group consisted of 10 typically developing children (6 boys and 4 girls) with typically-developing speech characteristics (as determined by a speech language pathologist) having an average age of $5.912\pm0.593$ years. The recorded stimuli was comprised of 24 sentences containing different target consonants~\cite{chapman2016americleft}. The Americleft database was used with an approval from Institutional Review Board (IRB) with IRB ID: STUDY00008224 and written consent was obtained from all participants.

\begin{figure}[tbh]
\centering
\includegraphics[height=50mm,width=1\linewidth]{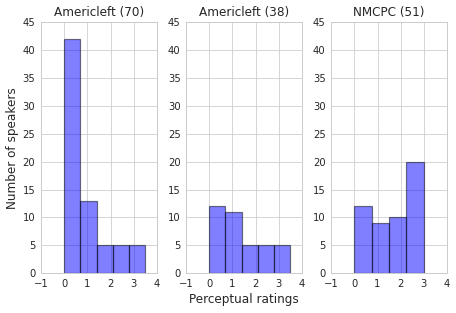}                                                      \caption{\label{hist} The distribution of the number of speakers over different ground-truth hypernasality levels. Histograms of weighted averaged ratings for (a) Americleft (70 speakers), (b) balanced Americleft (38 speakers) and NMCLP (51 speakers) databases.  }           \end{figure}                                                            
The hypernasality of the recorded speech samples was perceptually evaluated by 4 SLPs from the Americleft speech outcomes group (ASOG) according to a standardized protocol~\cite{chapman2016americleft}.  The speaker-level hypernasality was rated on the Americleft Speech Protocol scale on a 5-point scale (0-normal, 1-borderline, 2-mild, 3-moderate, 4-severe)~\cite{chapman2016americleft}.                                                        The Pearson correlation coefficient was computed between different pairs of raters to evaluate the inter-rater reliability.   The average inter-rater correlation coefficient was found to be $0.797\pm0.079$.  The ratings for all 4 SLPs were averaged to obtain a single ``ground-truth" rating per speaker~\cite{grimm2007primitives}. The histogram in Fig.~\ref{hist}(a) shows the distribution of hypernasality ratings for the 70 speakers from the Americleft database. It is clear from the figure that the database is skewed towards the normal (`0') end of the scale. We balance the original Americleft data by randomly removing a subset of speakers rated with normal hypernasality. The balanced Americleft database, Americleft(38), is comprised of 38 speakers and the histogram of the ground-truth ratings is shown in Fig.~\ref{hist}(b). The average inter-rater correlation for Americleft(38) was $0.776\pm0.068$.
                                                                     
In addition to hypernasality, the Americleft samples were evaluated for articulation errors.  The sentence stimuli were phonetically transcribed by the four Americleft raters using the International Phonetic Alphabet.  The number of active errors (glottal, pharyngeal, nasal fricatives, palatal, dental, lateral, double articulation errors) and passive errors (nasal substitutions, weak pressure consonants) were computed by the four SLP raters.  In the present work, the active and passive errors were reported against the target consonants present in the 24 sentence-level recordings.    Finally, for each speaker, the ratio between  the number of active errors and the total number of target consonants was used to compute the percentage of active errors. Similarly, for each speaker, we computed the percentage of passive errors.      
                                                                 
\subsection{\bf The New Mexico Cleft Palate Center database}
The New Mexico Cleft Palate Center (NMCPC) database is described in \cite{javid2020single}.  The database is comprised of speech samples from 10 controls (8 boys and 2 girls) and 41 children with CP (41 speakers (22 boys and 19 girls) with an average age of $9.2\pm3.3$ years. Each child was asked to repeat a random subset of sentences selected from a larger set of 76 sentences. The number of sentences per participant ranged from 7 to 69. The recorded samples were perceptually evaluated by 5 listeners on a continuous scale ranging from 0 to 3, where 0 stands for normal and 3 for severe hypernasality. These raters were not speech language pathologists, but they were speech processing experts that listened to the samples together after some self-training on how to evaluate hypernasality. The average inter-rater correlation of this database was $0.872\pm0.078$. The ratings of 5 raters were averaged to obtain  a single ``ground-truth" rating per speaker \cite{grimm2007primitives}.    A histogram of average ratings is shown in Fig.~\ref{hist}(c). This database is nicely balanced across the different hypernasality levels. The NMCPC database is a publicly available database that can be acquired upon request to  Dr. Luis Cuadros, New Mexico Cleft Palate Center.                                                                                                  \section{Methods}
\subsection{DNN Nasality Model}
Below we describe the development of the DNN nasality model, including its architecture and the training procedure. \\
\noindent{\bf DNN model architecture:}

\begin{figure}[tbh]
\vspace{-0.2cm}
\centering
\includegraphics[height=50mm,width=\linewidth]{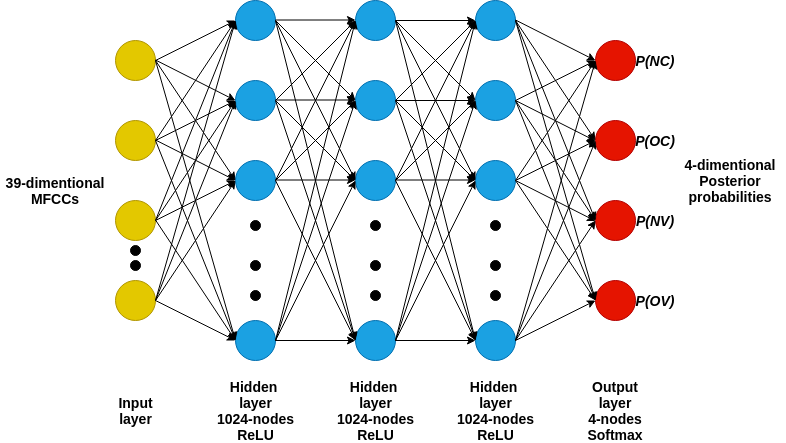}
\caption{\label{DNN} The architecture of the DNN nasality model. The model is a feed forward neural network consisting of a 39-dimensional input layer, three hidden layers with 1024 hidden neurons in each layer, and a 4-dimensional softmax output layer. The output layer yields posterior probabilities corresponding to nasal consonants (NC), oral consonants (OC), nasal vowels (NV), and oral vowels (OV).  } \end{figure}

\begin{figure}[tbh]
      \centering                  
      \includegraphics[height=50mm,width=\linewidth]{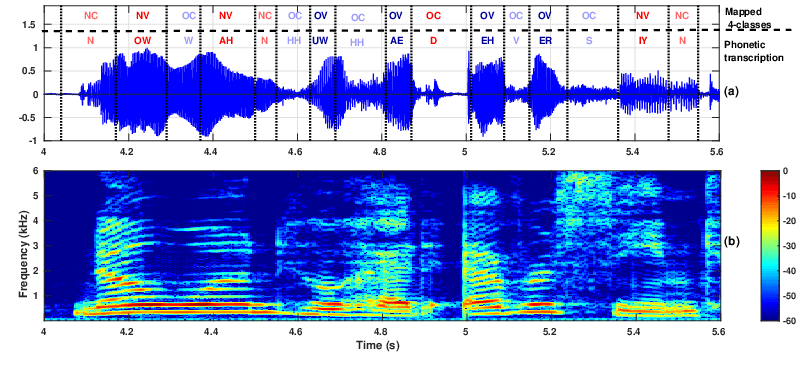}   
                  
      \caption{\label{intro_waveform}An illustration of the phoneme mapping procedure used for DNN training. (a) The speech waveform corresponding to the text `no one who had ever seen'  and (b) its spectrogram. Overlaid on the waveform, we show the transcription in English ARPABET format and the mappings to the four classes of interest, i,e. nasal consonant (NC), oral consonant (OC), nasalized vowel (NC), and oral vowel (OV).}           
     \end{figure}

The architecture of the DNN nasality model is shown in Fig.~\ref{DNN}. The model layer has an input layer with 39-nodes, corresponding to the 39-dimensional MFCCs input speech feature. The model is comprised of 3-hidden layers, where each layer has 1024 hidden neurons with rectified linear unit (ReLU) activation. The output layer consists of 4 softmax nodes, each interpreted as a posterior probability corresponding to nasal consonant (NC), oral consonant (OC), nasal vowel (NV), and oral consonant (OC) classes. 
\noindent{\bf  Training the DNN:}
First, the 100 hours of heatlhy speech recordings of Librispeech corpus and the corresponding orthographic transcriptions were passed though the Montreal forced-aligner~\cite{mcauliffe2017montreal} to align the speech acoustics to the transcript at the phoneme level. The segmented phonemes were grouped into the four classes of interest: nasal consonants (NC), oral consonants (OC), nasalized vowels (NV), and oral vowels (OV).  The NC group was formed by combining across nasal consonants (/N/, /M/, and /NG/); the OC group was formed by combining across oral consonants, including  plosives (/B/, /D/, /G/, /P/, /T/, /K/), fricatives (/Z/, /ZH/, /V/, /S/, /SH/, /F/, /H/), affricates (/JH/, /CH/), glides and liquids (/L/, /R/). The NV group was formed by combining thirty percent of the vowel segments that follow and precede a nasal consonant. The OV group was formed by combining across the remaining vowels segments, which were not surrounded by nasal consonants.    An example grouping of phonemes in a healthy speech sample is illustrated in Fig.~\ref{intro_waveform}. The speech waveform corresponding to the phrase ``no one who had ever seen'' and its spectrogram are shown in Fig.~\ref{intro_waveform}(a) and (b), respectively.  The  English phonemes in ARPABET encoded form, along with their time boundaries are marked on the speech waveform in Fig.~\ref{intro_waveform}(a). Based on the velopharyngeal activity, the phonemes are grouped into NC, OC, NV, and OV categories. In the example shown in Fig.~\ref{intro_waveform}, the nasal consonant (/N/) and the vowels (/OW/, /IY/) surrounding it are grouped into NC and NV classes, respectively. The oral consonants (/W/, /HH/, /D/, /V/, /S/) and the vowels (/UW/, /AE/, /EH/, /ER/) adjacent to them are grouped as OC and OV classes, respectively. 

The input speech to the DNN was sampled at a 16 kHz sampling rate, and short-time processed using a 20 ms Hamming window with 10 ms overlap. From each frame, 13 dimensional MFCCs, velocity ($\Delta$) and acceleration ($\Delta\Delta$) coefficients were computed. This 39-dimensional feature vector was the input to the  DNN nasality model; the label for each 20 ms frame corresponded to the category to which that frame belonged to. The classifier was trained to classify between the four phoneme categories described above. The error between the predicted and ground truth labels was computed using a categorical cross-entropy loss function. The ADAM optimizer was used to estimate the optimum parameters of the network. The network was trained for 25 epochs with a learning rate of 0.001. The MFCC features were computed using the Librosa package in Python %\footnote{$https://librosa.org/librosa/generated/librosa.feature.mfcc.html$} 
 and the DNN was implemented using the Keras 2.2.4 toolkit with a TensorFlow 1.13.1 backend. %\footnote{$https://keras.io/about/$}.

\begin{figure}[tbh]
\centering
\includegraphics[height=60mm,width=\linewidth]{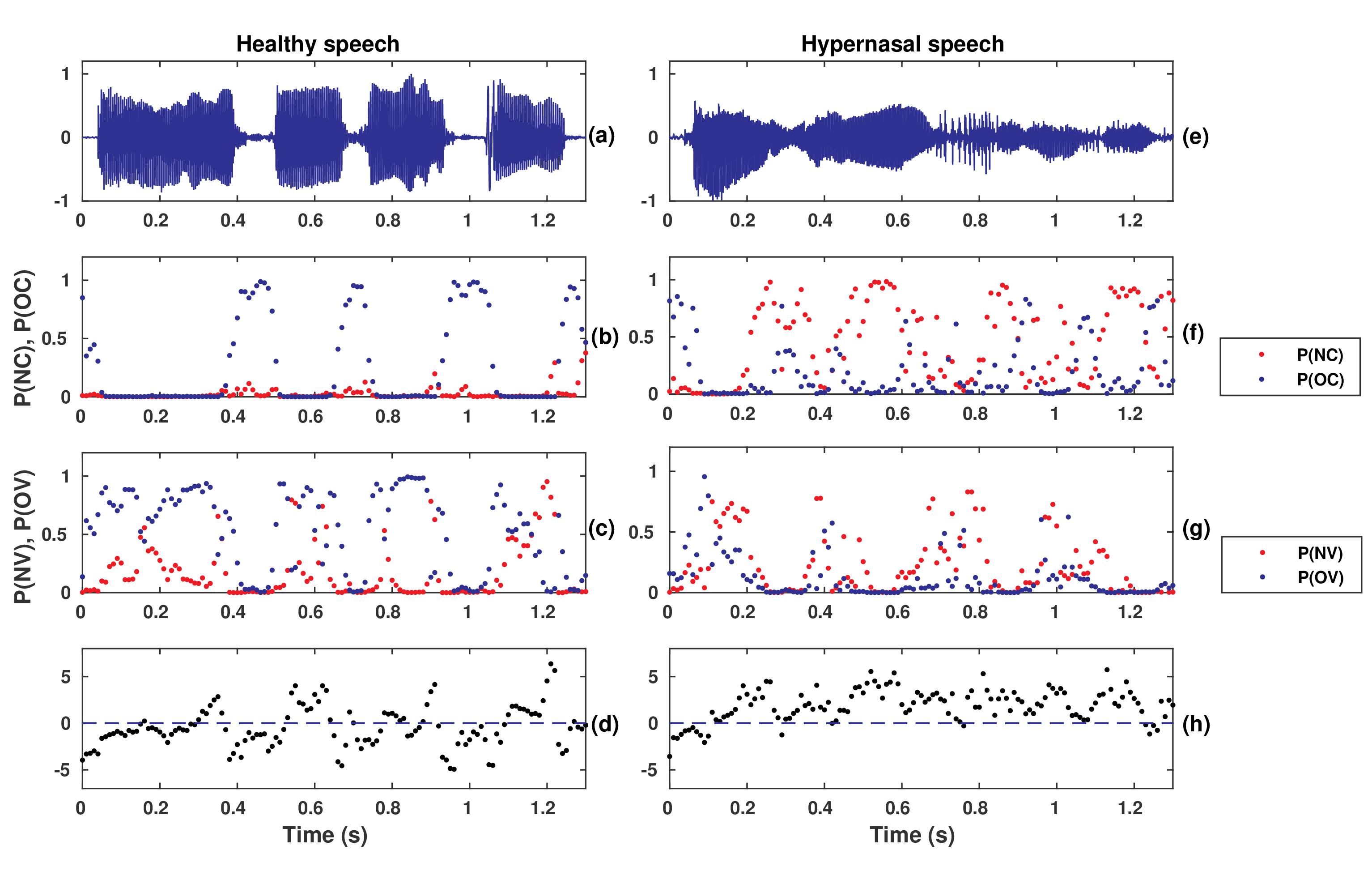}
\caption{\label{nasality_ratio} The four frame-wise DNN posteriors and corresponding frame-wise OHM for healthy and hypernasal speech: (a) the waveform of a control speech sample (``buy baby a bib"), (b)  $P(NC)$ and $P(OC)$, (c) $P(NV)$ and $P(OV)$, and (d) the OHM for the target sentence produced by a child from the control group;  (e)  the waveform of a CP speech sample (``buy baby a bib"), (f)  $P(NC)$ and $P(OC)$, (g) $P(NV)$ and $P(OV)$, and (h) OHM for target oral sentence produced by a participant with CP.  }
                                                                                 
\end{figure}

\noindent{\bf DNN posteriors as nasality features:}
                                                                    %Here describe how you compute the four DNN posteriors and explicitly state that we consider these as the nasality features (this will be a short section)
For the given input speech frame, the DNN results in 4 posterior probabilities corresponding to the NC, OC, NV, and OV classes. As described in Fig.~\ref{nasality_ratio}, increased values of $P(NC)$ and $P(NV)$ indicate the presence of nasalization in consonants and vowels, respectively.  Hence, we consider these posteriors as the nasality features and we used these to compute an objective measure of hypernasality.

\noindent{\bf Evaluating the OHM:}
 Evaluating the OHM requires the pre-trained DNN nasality model described above. The DNN was trained on adult speech, however we aim to use it to evaluate hypernasality in children's speech.  To compensate for the acoustic mismatch between children and adult speech, we used the pitch modification algorithm proposed by~\cite{shahnawazuddin2017effect}. The pitch modification pre-processing step lowers the pitch and speaking rate of children per the details in the paper. This same approach was used to improve the performance of speech recognition algorithms trained on adult speech and evaluated on children speech ~\cite{shahnawazuddin2017effect}. The pitch modification algorithm was implemented in MATLAB 2019a. 
                                                                    
 The pitch-modified speech signal was resampled at 16 kHz and short-time processed using a 20 ms Hamming window with 10 ms overlap. The frame-size and frame-shift were consistent with the parameters used during DNN nasality model training. As before, a 39-dimensional MFCC feature vector was computed and provided as input to the pre-trained DNN nasality model. The 4 DNN posterior probabilities were computed for each frame of children's speech.  
                                                                    
The DNN posteriors obtained for the pre-processed children's speech were used to compute the OHM. Let, $x_i$ be the feature vector corresponding to the MFCC input features for the $i^{th}$ frame; the DNN outputs probabilities corresponding to NC, OC,  NV, and OV  classes for that frame, i.e.,  $P(NC|x_i)$, $P(OC|x_i)$,  $P(NV|x_i)$, and $P(OV|x_i)$, respectively. Then the objective hypernasality  measure $OHM(x_i)$ for $i^{th}$ frame is computed as
       \begin{equation}
     OHM(x_i) =  max\left ( log (\frac{P(NC|x_i)}{P(OC|x_i)}) ,  log(\frac{P(NV|x_i)}{P(OV|x_i)})\right)
                    \label{n_x} 
   \end{equation}
   
In the above equation~\ref{n_x}, we compute the ratios of posterior probabilities of nasal to oral consonants and nasalized to oral vowels. If either ratio is larger than 1, it indicates the presence of a nasalized sound. The frame-level OHM, $OHM(x_i)$, is built by logarithmically transforming each ratio and taking the maximum across the two.

   \begin{table*}[tbh]
   \centering
     \caption{\label{Sentence_result} List of sentences in the Americleft database, target consonants, and the sentence-level correlation values to the OHM.   } 
   \ra{1.2}
   \resizebox{0.9\linewidth}{!}{
   \begin{tabular}{cclccccc}
   
   \hline\hline
   Sl. No & Target    & \multicolumn{1}{c}{Sentence} & r      & $p$-value & Category   & \begin{tabular}[c]{@{}c@{}}r \\ (Category)\end{tabular} & \begin{tabular}[c]{@{}c@{}}$p$-value\\ (Category)\end{tabular} \\ \hline
   1      & P         & Puppy will pull a rope       & 0.414  & $<0.05$   & Plosives   & 0.703                                                   & $<0.001$                                                       \\
   2      & B         & Buy baby a bib               & 0.568  & $<0.001$  &            &                                                         &                                                                \\
   3      & T         & Your turtle ate a hat        & 0.586  & $<0.001$  &            &                                                         &                                                                \\
   4      & D         & Do it today for dad          & 0.658  & $<0.001$  &            &                                                         &                                                                \\
   5      & K         & A cookie or a cake           & 0.568  & $<0.001$  &            &                                                         &                                                                \\
   6      & G         & Give aggie a hug             & 0.651  & $<0.001$  &            &                                                         &                                                                \\ \hline
   7      & F         & A fly fell off a leaf        & 0.715  & $<0.001$  & Fricatives & 0.793                                                   & $<0.001$                                                       \\
   8      & V         & I love every view            & 0.716  & $<0.001$  &            &                                                         &                                                                \\
   9      & TH        & Thirty-two teeth             & 0.616  & $<0.001$  &            &                                                         &                                                                \\
   10     & DH        & The other feather            & 0.63   & $<0.001$  &            &                                                         &                                                                \\
   11     & S         & Sissy saw sally race         & 0.691  & $<0.001$  &            &                                                         &                                                                \\
   12     & Z         & Zoey has roses               & 0.471 & $<0.05$   &            &                                                         &                                                                \\
   13     & SH        & She washed a dish            & 0.724  & $<0.001$  &            &                                                         &                                                                \\
   14     & S-cluster & I spy a starry sky           & 0.540  & $<0.001$  &            &                                                         &                                                                \\
   20     & H         & Hurry ahead harry            & 0.663  & $<0.001$  &            &                                                         &                                                                \\ \hline
   15     & CH        & Watch a choo-choo            & 0.715  & $<0.001$  & Affricates & 0.752                                                   & $<0.001$                                                       \\
   16     & J         & George saw gigi              & 0.698  & $<0.001$  &            &                                                         &                                                                \\ \hline
   17     & L         & Laura will yell              & 0.487 & $<0.05$   & Liquids    & 0.564                                                   & $<0.001$                                                       \\
   18     & R         & Ray will arrive early        & 0.507 & $<0.05$   &            &                                                         &                                                                \\ \hline
   19     & W         & We were away                 & 0.640   & $<0.001$  & Glides     & 0.640                                                   & $<0.001$                                                       \\ \hline
   21     & M         & Mom and amy are home         & 0.069  & 0.682     & Nasals     & 0.108                                                   & 0.518                                                          \\
   22     & N         & Anna knew no one             & -0.126 & 0.452     &            &                                                         &                                                                \\
   23     & NG        & We are hanging on            & 0.269  & 0.103     &            &                                                         &                                                                \\
   24     & N, M, NG  & We ran a long mile           & 0.162  & 0.331     &            &                                                         &                                                                \\ \hline\hline
   \end{tabular}
   }
   \end{table*}                                                                    
For clarity, it is useful to demonstrate the OHM with two examples.
 The speech waveform, frame-wise DNN nasality posterior features, and the OHM contours from a control sample and a CP sample are plotted in Fig.~\ref{nasality_ratio}(a)-(h). The target sentence is ``buy baby a bib". Since the sentence does not contain any nasal consonants, no nasal cues are expected in the speech from the control group; this is consistent with panels (b) and (c) where $P(OC)>P(NC)$ and $P(OV)>P(NV)$ for the speech from the control group. For the case of hypernasal speech from the CP group in panels (f) and (g), we see that $P(NC)>P(OC)$ and $P(NV)>P(OV)$. Although the target text does not contain any nasal consonants, large $P(NC)$ and $P(NV)$ values indicate the presence of abnormal nasal resonances in the CP speech indicative of hypernasality. As expected, the OHM measure obtained by combining $P(NC),~P(OC),~P(NV),$ and $P(OV)$ indicates relatively higher values for hypernasal speech (Fig.~\ref{nasality_ratio}(d) than for healthy speech (Fig.~\ref{nasality_ratio}(h)).
                                                                                                
 The frame-level OHM scores ($OHM(x_i)$) were averaged over all the  frames of a given utterance to obtain sentence-level OHM scores. Similarly, sentence-level OHM scores were averaged over all utterances spoken by the same speaker to obtain speaker-level OHM scores.                

\section{Experiments and Results}
 The experiments were conducted using the Americleft database to the evaluate the sentence and speaker-level performance of the OHM, the robustness of OHM to active errors, the sensitivity of OHM, and the internal reliability of OHM. The external validity of the OHM was evaluated using NMCPC database. Further, a comparison of the OHM with a representative supervised learning method based on DNN regression was also conducted. The details of the experiments and the results are presented in the following subsections.

\subsection{Validation of sentence-level OHM scores}
 
The frame-level results in Fig.~\ref{nasality_ratio} are averaged over the entire duration of the utterance to generate a sentence-level OHM score. The correlation between the sentence-level OHM and the speaker-level perceptual ratings, the ground-truth rating obtained from Americleft-trained SLPs, was evaluated using Pearson's correlation coefficient ($r$).  In Table~\ref{Sentence_result} we list the sentences from the Americleft database, grouped by target consonant category; for each sentence, we also list the correlation between the sentence-level OHM and the perceptual hypernasality level. As expected, oral sentences, i.e., the sentences containing oral consonants (plosives, fricatives, affricates, liquids, and glides) showed a high correlation with the perceptual ratings; whereas the OHM calculated from nasal sentences reveals a low correlation. This makes sense as the OHM demonstrates a ceiling effect for nasal sentences since it is expected  that they are nasalized.

\subsection{Validation of speaker-level OHM scores}

We compute the speaker-level OHM scores by averaging across all oral sentences produced by a speaker. The average Pearson correlation between the speaker-level OHM with each rater's perceptual ratings \emph{and} the average inter-rater Pearson correlation are shown in Table~\ref{Americleft-res}. 

\begin{figure}[tbh]
             \vspace{-0.2cm}
                 \centering
                      
   \includegraphics[height=60mm,width=0.9\linewidth]{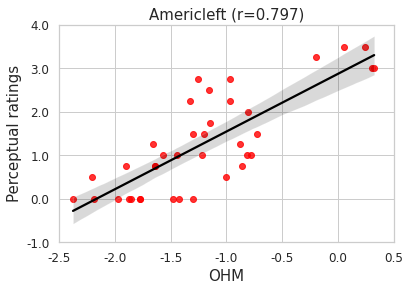}
    \centering           
 \caption{\label{result} A scatter plot of speaker-level OHM scores vs. ground-truth perceptual ratings for the Americleft database.  }
 \end{figure}

For finer-grained analysis, we compare the OHM with the ground-truth rating obtained by averaging the clinical ratings from the 4 raters. Fig.~\ref{result} shows the relationship between the speaker-level OHM and perceptual ratings for the Americleft database. The OHM shows a significant correlation ($r=0.797,~p<0.001$) with the ground truth perceptual ratings. A scatter plot of sample-level data is shown in Fig. \ref{result}.

\begin{table}[tbh]
   \centering
   	 \caption{\label{Americleft-res} A comparison of the average pairwise correlation between Americleft-trained raters and the average correlation between the OHM and each rater. } 
   \ra{1.4}
  \resizebox{0.5\linewidth}{!}{
\begin{tabular}{lc}
\hline\hline
             & Mean$\pm$std.   \\ \hline 
Inter-rater correlation    & 0.776$\pm$0.068 \\
OHM-rater correlation      & 0.733$\pm$0.051  \\
\hline\hline
\end{tabular}
}
\end{table}

\subsection{Robustness to active articulation errors}

  We analyzed the effect of articulation errors on the estimated OHM scores.  Articulation errors in CP cases are broadly categorized into active and passive errors~\cite{henningsson2008universal, harding1998active}. The percentage (\%) of active articulation errors (PAAE) and percentage (\%) of passive articulation errors (PPAE) were computed at the speaker-level using the 24 Americleft sentences. A correlation between the Americleft ratings and the PAAE and PPAE was evaluated, as was a correlation between the OHM scores and PAAE and PPAE. 

\begin{figure}[tbh]

\centering
                    
\includegraphics[height=55mm,width=\linewidth]{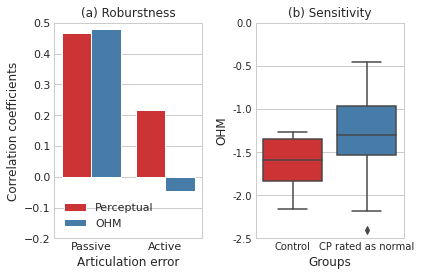}
\caption{\label{fig-rel} (a) Correlation of OHM scores and perceptual hypernasality scores with respect to active and passive articulation errors.  (b) OHM scores for the control group and for children with CP rated as having normal hypernasality in the Americleft database. }             
\end{figure}

The bar plot in Fig.~\ref{fig-rel}(a) shows the correlation of the perceptual hypernasality ratings with respect to the PAAE and PPAE. The Americleft ratings showed a moderate correlation ($r=0.460,~p<0.001$) with respect to PPE. Passive errors include nasalized consonants, which carry nasal cues, therefore, it is expected that the presence of passive errors increases the severity of perceived hypernasality. In fact, the perception of nasalized consonants was considered an important criterion in developing  the hypernasality rating scale in~\cite{henningsson2008universal}.  Nasal resonances are not evident in active errors, such as glottal stops, pharyngeal stops, and nasal fricatives. Hence, the perceptual ratings showed a low correlation ($r=0.218,~p=.189$) with the Americleft ratings for the active errors. 

Similar to the hypernasality ratings, the OHM scores also showed a moderate correlation ($r=0.481,~p>0.05$) with PPAE and a low correlation ($r=-0.048,~p=0.773$) with respect to PAAE.  However, when compared to perceptual ratings, the OHM showed a relatively lower correlation for active errors and higher for passive errors. These results provide additional evidence that the OHM captures the nasal cues present in the speech signal and is robust to co-existing active articulation errors.

\subsection{Evaluating the sensitivity of the OHM} 
In our analysis, we considered a balanced set of 38 speakers in Americleft database  to evaluate the sentence-level and speaker-level OHM scores relative to the perceptual ratings. Additionally, we analyzed the OHM for the remaining 32  `CP speakers rated as normal' (no hypernasality) and compared them with the control group. Here, the  `CP rated as normal' corresponds to speakers whose SLP hypernasality rating was considered normal.  Fig.~\ref{fig-rel}(b) shows the range of the OHM for the two groups. The OHM scores of speakers with CP rated `0' were greater than that of  controls. A  $t$-test reveals a statistically significant difference between the two groups ($t=-2.899$, $p<0.05$).  This result may indicate the presence of very mild hypernasality in the CP group not detected by the SLPs.

\subsection{Assessing the internal reliability of the OHM}  

\begin{figure}[tbh]
\centering
\includegraphics[height=60mm,width=0.95\linewidth]{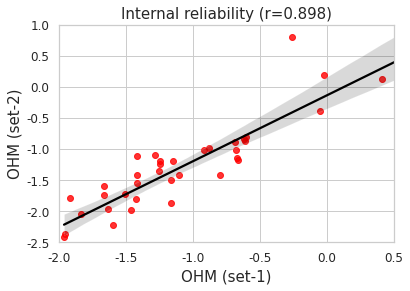}
                 
\caption{\label{fig-testre}Analysis of internal reliability. A scatterplot of speaker-level OHM scores computed for two independent sets of sentences. }
                  
\end{figure}

To evaluate the internal reliability of OHM scores, we grouped the 20 oral sentences from the Americleft database into set-1 and set-2, where set-1 contains the first 10 sentences and the remaining 10 formed set-2. We computed speaker-level OHM scores for  set-1 and set-2 data. Fig.~\ref{fig-testre} shows the scatter plot of OHM scores for  set-1 vs. set-2. The OHM scores of set-1 are significantly correlated ($r=0.898, ~p<0.001$) with that of set-2. 
                                      
\subsection{Assessing the external validity of the OHM}
We also evaluated the performance of the OHM on the NMCPC database in order to evaluate how well the OHM generalizes to data collected in other studies and evaluated by other SLPs. The hypernasality level of the NMCPC speech samples was evaluated by 5 SLPs. The correlation of the OHM with respect to each rater and the inter-rater correlation are shown in Table~\ref{NMCPC}. The average correlation of the OHM vs. individual raters was found equal to $r=0.695,p<0.001$ whereas the inter-rater correlation was $r=0.872, p<0.001$.

\begin{figure}[tbh]
               \vspace{-0.2cm}
                   \centering
                        
     \includegraphics[height=60mm,width=0.95 \linewidth]{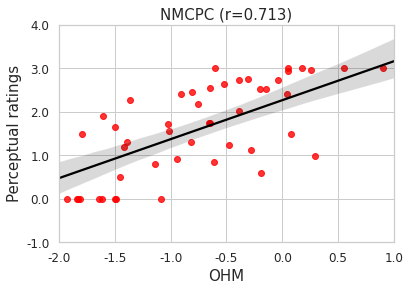}
                 
   \caption{\label{plot-NMCPC}  A scatter plot of speaker-level OHM scores vs. ground-truth perceptual ratings for the NMCPC database. }

   \end{figure}

 The scatter plot of the speaker-level OHM vs. the average of the 5 clinical ratings is shown in Fig.~\ref{plot-NMCPC}.   The OHM showed a significant correlation ($r=0.713, ~p<0.001$) with the average hypernasality ratings provided by the SLPs.
  
  \begin{table}[tbh]
   \centering
   	 \caption{\label{NMCPC}  A comparison of the average pairwise correlation between NMCPC raters and the average correlation between the OHM and each rater.  } 
   \ra{1.4}
  \resizebox{0.6\linewidth}{!}{
  \begin{tabular}{lc}
  \hline\hline
              & Mean$\pm$std.        \\ \hline
  Inter-rater correlation     & 0.872$\pm$0.078 \\
  OHM-rater correlation         & 0.695$\pm$0.016 \\ \hline\hline
  \end{tabular}
  }
  \end{table}

\subsection{Comparison with a fully supervised approach}

 Conventionally,  automatic evaluation of hypernasality involves the supervised training of models like SVMs~\cite{golabbakhsh2017automatic,orozco2013nonlinear}, artificial neural networks~\cite{wang2019automatic}, and recurrent neural networks~\cite{wang2019hypernasalitynet} for the binary classification of healthy and hypernasal speech samples. In all of these existing approaches, the supervised training of models was carried out by using a perceptually labeled CP speech database.    In the proposed approach, the OHM was computed using the nasality DNN, whose parameters were estimated by using only the healthy speech samples. The existing deep learning-based implementations~\cite{wang2019automatic, wang2019hypernasalitynet} were  aimed for binary classification from the segmented vowels. We found only one DNN-based work in the literature~\cite{vikram2018estimation}, which predicts hypernasality severity from the connected speech samples. In this work~\cite{vikram2018estimation}, the DNN was training directly on a labeled CP speech database.    To compare the OHM with a conventional supervised approaches, we implemented a DNN regressor, which was directly trained on the MFCC features extracted from the speech samples and the perceptual ratings of the Americleft database.
 
  We used only oral sentences to train and test the DNN regressor using leave-one-speaker-out (LOSO) cross-validation. The sample size of the Americleft database (20 oral sentences x 38 speakers=760) is very small to train a DNN. To address this issue, we used data augmentation using (a) addition of noise: white, babble, and factory noise with 5, 10, 15, and 20 dB SNRs, (b) speaking rate modification using the factors 0.8, 0.9, 1.1, and 1.2, (c) vocal tract length perturbation (VTLP) using the perturbation factors 0.9, 0.95, 1.05, and 1.1~\cite{lo2020ntnu}. After the data augmentation, the sample-size of the database was increased from 760 to 9120 sentence-level recordings. 

 The 39-dimensional MFCC features extracted using 20 ms Hamming windowed speech frames with a shift of 10 ms were fed to a feed-forward DNN regressor. The architecture details of DNN regression are as follows: 39 input nodes, three hidden layers with 512 neurons with ReLU activation, and 1 output node with a linear activation function. The error between the predicted outputs and ground truth labels was estimated using the mean squared error (MSE) loss function. The ADAM optimizer was used to estimate the optimum parameters of the network. The network was trained for 25 epochs, with a learning rate of 0.001. 	

The MFCCs were computed at the frame-level, but the ground truth was available at the speaker-level. During training we assigned the speaker-level hypernasality ratings to every frame-level feature vector belonging to that particular speaker. In the testing phase, speaker-wise averaging of DNN outputs was carried out to get a single score per speaker.  The performance of the DNN regressor was evaluated using the leave-one-speaker-out (LOSO) cross-validation criteria. In LOSO cross-validation, the samples of all speakers except one speaker were used to train DNN and the remaining one's sample was considered for the testing. Note that the augmented samples were used only during the training phase, while testing we used only the original samples.

	 \begin{table}[tbh]
	  \caption{\label{Comp-res}  Comparison between OHM and DNN regressor.  } 
	 \centering
	 \ra{1.5}
	 
	 \resizebox{1\linewidth}{!}{
	 \begin{tabular}{lll}
	 \hline\hline
	 \multirow{2}{*}{Database} & \multicolumn{2}{c}{Approach}                                                                                                                                  \\ \cline{2-3} 
	                           & \multicolumn{1}{c}{\begin{tabular}[c]{@{}c@{}}OHM\end{tabular}} & \multicolumn{1}{c}{\begin{tabular}[c]{@{}c@{}}DNN Regressor\end{tabular}} \\ \hline
	 Americleft                &  $r=0.797$, $p<0.001$                                                 & $r=0.524$, $p<0.05$                                                           \\
	 NMCPC                     & $r=0.713$, $p<0.001$                                                 & $r=0.301$, $p=0.032$                                                                     \\ \hline\hline
	 \end{tabular}
	 }
	 \end{table}
	   
The correlation coefficient computed between predicted scores by DNN-regressor  and the perceptual ratings is  shown in Table~\ref{Comp-res}. The correlation found to statistically significant ($r=0.524,~p<0.05$), but it is well below that of the OHM. The DNN regressor trained on Americleft samples was used to evaluate the hypernasality in NMCPC samples and the results are presented in Table. The predicted scores for NMCPC database showed a weak correlation ($r=0.301, ~p=0.032$) with the perceptual ratings. These results indicate  overfitting effect. The  OHM scores resulted in a strong  correlation ($p<0.001$) for both the databases and these results empirically show that the OHM was robust to a variety of recording conditions, sentence contexts, and gold-standard perceptual ratings. 

 \section{Discussion}
 In this work,  we introduced an objective measure of hypernasality based on a DNN nasality model trained on healthy speakers with no clinical labels.  First, we modeled the acoustic cues related to an open velum (nasalized) from healthy speech by training a DNN classifier to classify between NC, OC, NV, and OV classes. This pre-trained DNN on healthy  speech samples was used to characterize the presence of abnormal nasal resonances in the speech of children with CP.  The OHM was computed at the level of a speech frame and aggregated into sentence-level and speaker-level scores.  The OHM is found to have several advantages over the existing hypernasality evaluation methods as the approach does not require a labeled clinical database and orthographic transcriptions.

   \vspace{0.2cm}
     \noindent  {\bf Comparison with existing hypernasality evaluation approaches:} 
 Most of the existing approaches for the automatic hypernasality evaluation in children with CP were aimed at the classification of normal and hypernasal speech~\cite{wang2019automatic,golabbakhsh2017automatic} or multi-class classification (normal, mild, moderate, and severe)~\cite{dubey2019detection, javid2020single}.  Since these supervised machine learning models were directly trained on CP speech samples and the perceptual hypernasality ratings, their performance critically depends on the availability of labeled hypernasality speech database.  To analyze the risk of overfitting in supervised training in the existing approaches and the advantages of the OHM, we implemented and evaluated a hypernasality estimation method based on DNN regression. The regressor was trained directly on the speech samples and ratings of the Americleft database. Even though we increased the database size via data augmentation, the DNN regressor's was well below that of the OHM. The availability of a perceptually labeled database became a practical limitation in clinical speech research due to the limited number of subjects and trained SLPs. The OHM bypasses the need for clinical labels as the scores are estimated from a DNN trained on a publicly available large healthy speech corpus. We only use the clinical data to evaluate the OHM with the perceptual ratings.

  The DNN regressor trained on the Americleft database was also tested on the NMCPC database with poor results. Since the model was trained for the Americleft samples and ratings, the model overfit to the training examples and showed a  lack of generalization ability on another database. In contrast, the results in  Table~\ref{Comp-res} revealed that the OHM is robust to differences in speech samples and rating scales. This is evidenced by the fact that the perceived hypernasality of the Americleft samples was rated on a 5-point scale, whereas the NMCPC data was rated on a 4-point scale. Since the training of the DNN does not use any perceptual ratings, the OHM itself is not sensitive to the rating scales. In fact, our results empirically show that the OHM was robust to a variety of recording conditions, sentence contexts, and gold-standard perceptual ratings. 
  
  Another important advantage of OHM is that the approach does not require phonetic segmentation. Most of the automated hypernasality evaluation methods rely on the phonetic segmentation, where segmentation was performed either by manual labeling or force-alignment using orthographic transcriptions. The OHM was computed directly on the connected speech samples and this makes the evaluation process simple and faster as manual labeling and transcribing speech is a time-consuming process.

  \vspace{0.2cm}
   \noindent   {\bf The role of the speech stimuli for estimating hypernasality:} The choice of stimuli or target sentence plays an important role in the assessment of hypernasality.  In the case of healthy speakers, nasal resonances are inherently present during the production of nasal sentences and completely absent in the case of oral sentences. The presence of nasal resonances during the production of oral sentences is considered to be abnormal, which indicates the presence of hypernasality~\cite{henningsson2008universal}. Sentence-wise correlation values listed in  Table~\ref{Sentence_result} revealed that the proposed measure yields a good correlation for the  oral sentences and poor correlation for nasal sentences. For both CP subjects and controls, the acoustic energy passes through the nasal tract while producing nasal sentence; hence, it is difficult to evaluate hypernasality using nasal sentences. These results indicate that to reliably compute the speaker-level OHM, the target speech samples should only contain oral consonants. These results closely match with the  perceptual assessment guidelines mentioned in~\cite{henningsson2008universal}, where only the sentences with oral consonants were suggested for assessing hypernasality. 
  
  High-pressure and low-pressure consonants play an important role in clinical evaluation of hypernasality~\cite{henningsson2008universal, kummer1996evaluation}.  
  Pressure-sensitive or high-pressure consonants (plosives: /P/, /T/, /K/, /B/, /D/, /G/,  fricatives: /S/, /F/, /SH/, /Z/, /V/, and affricates: /CH/, /JH/, /TH/, /DH/) require adequate intraoral pressure, which is developed by the closure of oral and nasal tracts. Whereas,  low-pressure consonants (glides: /Y/,  /W/ and liquids: /L/, /R/)  do not require high intra-oral pressure.   Since the loss of airflow in patients with CP and VPD affects the ability to build up and maintain intra-oral air pressure, the production of  high pressure consonants is severely affected. The substitution of nasal consonants for  target high pressure consonants is most commonly reported in speakers with CP because of the escape of air through the incompletely closed VP port . Therefore, in clinical settings, target speech  containing high-pressure oral consonants is highly recommended for the perceptual assessment of hypernasality~\cite{henningsson2008universal, kummer1996evaluation}. Our results are consistent with this as the OHM shows higher correlation  for sentences containing high-pressure consonants (plosives, fricatives, and affricates) than those with low-pressure consonants.

 \vspace{0.2cm}
  \noindent {\bf Active and passive articulation errors in assessment of hypernasality:}  The articulation errors in speakers with CP are widely divided into active and passive articulation errors. In the case of passive articulation errors,  the speaker tries to retain the place of articulation of the target phoneme but the air escapes through the velopharyngeal port. Hence, the consonant is perceived to be weak or nasalized. In the case of severe VPD, the target consonant is completely replaced by a nasal consonant (/T/, /D/$->$/N/, /P/, /B/$->$/M/) ~\cite{henningsson2008universal}. In the case of active errors (also known as compensatory errors), a speaker attempts to compensate for the effect of nasalization by  shifting  the location of articulatory constriction. The shift in the place of articulation may be within the oral or non-oral cavity. In many cases, the presence of VPD leads to a shift towards glottal and pharyngeal regions ~\cite{henningsson2008universal}. Although active and passive articulation errors are produced as a consequence of VPD, their contribution in the perception of hypernasality is different.  The passive errors contain nasal cues and hence, their presence contributes to the perception of hypernasality. In contrast, the active articulation errors, such as glottal and/or pharyngeal substitutions, do not carry nasal cues; therefore they do not contribute to the perception of hypernasality. However, the presence of active errors can create variability in perceptual evaluation as it is difficult to perceptually decouple active articulation errors from the presence of hypernasality. As shown in Fig. \ref{fig-rel}(a), the OHM scores do not have this bias as they show no correlation with the active errors. This result provides evidence that OHM scores capture only the nasal cues present in the speech signal and are robust to co-existing active errors in CP speech.

  \vspace{0.2cm}
  \noindent {\bf Differences in OHM performance on Americleft and NMCPC databases:} 
   For the Americleft database, the inter-rater reliabilities are on average a little higher than the  OHM-to-rater reliabilities, but they are within 1 standard deviation of each other. For the NMCPC database inter-rater reliabilities are significantly greater than the OHM-to-rater reliabilities.  Also, the OHM scores showed relatively higher correlation with the averaged perceptual ratings ($r=0.797$), when compared to the NMCPC database ($r=0.713$).    The reasons for this difference in the OHM's performance are  multifold. The  NMCPC database is not balanced in terms of the number of sentences per speaker. This imbalanced nature of NMCPC database affects the estimation of speaker-level OHM, where the speaker-level OHM was computed by averaging the sentence-level scores. The average signal-to-noise ratio (SNR) of the Americleft samples was found equal to 23.88$\pm$9.10 dB. Whereas the speech samples of NMCPC database are noisier with an SNR of  (15.95$\pm$8.10 dB). The raters of Americleft database were trained under the Americleft speech project and followed the same protocol~\cite{chapman2016americleft}. In the Americleft speech project, a group of SLPs with extensive expertise in the evaluation and treatment of children with CP developed a standard hypernasality rating scale and protocols for speech evaluation. The OHM showed good agreement with ratings that followed these guidelines, but not with the NMCPC raters  who were not trained in the Americleft standard protocols. Moreover, the NMCPC raters were speech researchers, but they were not SLPs and did not have clinical expertise in the domain of CP speech assessment. However, it is interesting that the NMCPC raters have a much higher inter-rater reliability than the Americleft ratings, well above what has been reported in the literature previously~\cite{chapman2016americleft}. We posit that this is because the raters discussed their ratings while evaluating the speech samples, reaching consensus in some cases.

\section{ Summary and future work}
In this work,  we introduced an objective measure of hypernasality based on a DNN nasality model trained on healthy speakers with no clinical labels.  First, we modeled the acoustic cues related to an open velum (nasalized) from healthy speech by training a DNN classifier to classify among NC, OC, NV, and OV classes. This pre-trained DNN on healthy  speech samples was used to characterize the presence of abnormal nasal resonances in the speech of children with CP.  The OHM was computed at the level of a speech frame and aggregated into sentence-level and speaker-level scores.
  
    In the current implementation of the OHM, the model is completely tuned and implemented using only the healthy speech database. Future work can focus on further refinement of the implementation by using a corpus of cleft speech to tune model parameters, or perhaps to change it to a supervised model by using linear regression across different sentences to produce speaker-level hypernasality scores.  The present work uses a pitch modification algorithm to compensate for the acoustic mismatch between adult and children's speech. However,  better speaker adaptation methods such as identity vector (i-vector) and transfer learning approaches can be explored to improve the system's performance. Another limitation of the current work is that the algorithm assumes the presence of information related to nasality over the entire duration of utterance and simply averages the frame-level OHM scores over an entire utterance.   However, depending on the severity level, the hypernasality information may be distributed unevenly over different phonemes.
     Therefore, instead of a simple averaging operation, recurrent neural networks and attention models can be used to capture  unevenly distributed nasality information.
     
    Furthermore, hypernasality is not only specific to CP speech. It can also be present in speech from individuals with neurological disorders, such as Huntington's and Parkinson's disease.  Therefore, future work can focus on extending the validation of this model for evaluating  hypernasality in dysarthric speech.

\section{ Acknowledgements}
Authors would like to thank Dr. Luis Cuadros, New Mexico Cleft Palate Center and Dr. Anil Kumar Vappula, IIIT Hyderabad, India for providing  NMCPC database. Also, authors would like to thank Dr. Nagaraj Adiga, a co-author of the work~\cite{shahnawazuddin2017effect} for sharing the pitch modification program.  %This work is funded in part by NIH grant NIDCR DE026252.
 \bibliographystyle{IEEEtran}
 \bibliography{refs.bib}
\end{document}